\documentclass[11pt]{article}
	
	\newcommand{\blind}{0}
	
    \makeatletter
    \renewcommand\section{\@startsection {section}{1}{\z@}%
                                       {-3.5ex \@plus -1ex \@minus -.2ex}%
                                       {2.3ex \@plus.2ex}%
                                       {\normalfont\fontfamily{phv}\fontsize{16}{19}\bfseries}}
    \renewcommand\subsection{\@startsection{subsection}{2}{\z@}%
                                         {-3.25ex\@plus -1ex \@minus -.2ex}%
                                         {1.5ex \@plus .2ex}%
                                         {\normalfont\fontfamily{phv}\fontsize{14}{17}\bfseries}}
    \renewcommand\subsubsection{\@startsection{subsubsection}{3}{\z@}%
                                        {-3.25ex\@plus -1ex \@minus -.2ex}%
                                         {1.5ex \@plus .2ex}%
                                         {\normalfont\normalsize\fontfamily{phv}\fontsize{14}{17}\selectfont}}
    \makeatother
	\usepackage{amsmath}
	\usepackage{graphicx}
	\usepackage{enumerate}
	\usepackage{xcolor}
	\usepackage{natbib} %comment out if you do not have the package
	\usepackage{url} % not crucial - just used below for the URL
        \usepackage{booktabs}
\begin{document}
		
			%%%%%%%%%%%%%%%%%%%%%%%%%%%%%%%%%%%%%%%%%%%%%%%%%%%%%%%%%%%%%%%%%%%%%%%%%%%%%%
		\def\spacingset#1{\renewcommand{\baselinestretch}%
			{#1}\small\normalsize} \spacingset{1}
		%%%%%%%%%%%%%%%%%%%%%%%%%%%%%%%%%%%%%%%%%%%%%%%%%%%%%%%%%%%%%%%%%%%%%%%%%%%%%%
		
		\if0\blind
		{
			\title{\bf Libraries, Integrations and Hubs for Decentralized AI using IPFS}
			\author{Richard Blythman$^{1}$ \quad Mohamed Arshath$^{1}$ \quad  Jakub Smékal$^{1}$ \\ Hithesh Shaji$^{1}$ \quad Sal Vivona$^{1}$ \quad Tyrone Dunmore$^{1}$ \\\\
			$^1$ Algovera.ai  }
			\date{}
			\maketitle
		} \fi
  
		\if1\blind
		{

            \title{\bf \emph{IISE Transactions} \LaTeX \ Template}
			\author{Author information is purposely removed for double-blind review}
			
\bigskip
			\bigskip
			\bigskip
			\begin{center}
				{\LARGE\bf \emph{IISE Transactions} \LaTeX \ Template}
			\end{center}
			\medskip
		} \fi
		\bigskip
		
	\begin{abstract}
AI requires heavy amounts of storage and compute. As a result, AI developers are regular users of centralised cloud services such as AWS, GCP and Azure, compute environments such as Jupyter and Colab notebooks, and AI Hubs such as HuggingFace and ActiveLoop. There services are associated with certain benefits and limitations that stem from the underlying infrastructure and governance systems with which they are built. These limitations include high costs, lack of monetization and reward, lack of control and difficulty of reproducibility. At the same time, there are few libraries that allow data scientists to interact with decentralised storage in the language that data scientists are used to, and few hubs where they can discover and interact with AI assets. In this report, we explore the potential of decentralized technologies - such as Web3 wallets, peer-to-peer marketplaces, decentralized storage (IPFS and Filecoin) and compute, and DAOs - to address some of the above limitations. We showcase some of the libraries and integrations that we have built to tackle these issues, as well as a proof of concept of a decentralized AI Hub app, that all use IPFS as a core infrastructural component. 
	\end{abstract}
			
	\noindent%

	%\newpage
	\spacingset{1.5} % DON'T change the spacing!

\newcommand{\RB}[1]{{\color{blue}{\bf RB: }{#1}}}

\section{Introduction} \label{s:intro}

The field of deep learning is powered by assets such as datasets, models and software, which require a powerful underlying infrastructure consisting of components like storage and compute. \cite{schwartz2020green} discuss how the expense associated with the 300,000x increase in compute requirements from 2012 to 2018 raises the barrier for participation in AI research (while also being environmentally unfriendly). Recently, there has been a trend towards open source software in AI, which has made significant contributions to research and applications (\cite{langenkamp2022open}). For example, the majority of models are implemented in open source libraries such as TensorFlow and PyTorch developed in the open source language Python. 

An AI Hub is a platform that allows data scientists, engineers and other stakeholders to discover, share and collaborate on AI assets such as code, datasets, models, apps, notebooks, pipelines and other software. AI Hubs have made significant contributions to the democratization of state-of-the-art research. For example, source code is commonly made available through hubs such as GitHub and HuggingFace Hub. Other assets such as datasets and pre-trained model weights are often open sourced through hubs such as Kaggle, HuggingFace and ActiveLoop Hub. As a result, data scientists are regular users of AI Hubs to provide a place for assets to be stored, shared and further developed. 

%Problems with existing assets
Despite much progress, many assets remain inaccessible to data scientists outside large organization. Datasets like JFT-300M (\cite{sun2017revisiting}) by Google and models like GPT-3 (\cite{brown2020language}) remain siloed for competitive reasons, while other companies and universities may lack the technical know-how to share assets. Furthermore, open assets can be scattered across many apps and websites, and requires the user to follow a lengthy tutorial for download, setup and processing. The assets are often isolated from compute environments and not well integrated with workflows, making reproducibility difficult. 

%Problems with existing hubs
While the assets themselves may be open source, the platform itself is almost always closed source and governed by a centralized entity. The trade-offs that have been made may arise from the availability and maturity of technologies and governance structures with which they are built. Today's AI Hubs tend to rely exclusively on centralised cloud services such as AWS, GCP, and Azure, and the high expense of these services is often passed on to the user. Cloud infrastructure is a critical component for training large scale models, and the computational cost for re-training or fine-tuning of open source models has become prohibitive and created a barrier-to-entry. Furthermore, the platform ultimately controls the accessibility of uploaded assets, and act as gatekeepers to which assets are allowed to exist on the respective platform. Finally, the platform can also monetize the network effects of user contributions without sharing in the rewards (\cite{marketplace}). 

%Other steps of the pipeline
Data science collaboration is complex and involves many stakeholders and stages (\cite{zhang2020data}). Model training is just one step within a pipeline, that also includes a number of other steps that make up the data science workflow. Many workflows are closed source with large companies building their own proprietary data pipelines (\cite{uvarov2022data}). \cite{langenkamp2022open} expect that data-centric tooling is the next frontier for AI research, although they note that the incentives for open sourcing these tools  - such as competitive differentiation - are different. Machine learning tooling for the rest of the pipeline is fragmented and suffers a tragedy of the commons.

%Potential solutions through decentralized infrastructure
Decentralized technologies such as peer-to-peer storage, compute and marketplaces, machine learning frameworks and decentralized autonomous organizations (DAOs) present opportunities for tackling the above challenges. We explore the benefits offered by these technologies to address some of the above issues. We propose a combination of a number of these technologies to offer an alternative value proposition to existing solutions. Metahub\footnote{https://metahub.algovera.ai/} is a decentralized, permissionless and censorship-resistant platform for data scientists, engineers, domain experts and users to build AI systems. 
% Metahub is composed of three layers, made up of infrastructure, modules and apps. The individual components within a layer are modular and interoperable, and can be composed to provide solutions to particular use cases. 
Algovera is a community that is coming together to abide by certain standards when creating and working with digital objects and building decentralized AI applications. The community agrees to recognize each others property rights according to a certain set of rules, and achieve consensus on ethical, safety, and alignment considerations. This arrangement can ultimately be coded in software, with collective governance to update the standards. This structure of coordination facilitates a pluralist approach to AI (\cite{siddarth2021how}). 

% With Metahub, users can store, share, and search for assets stored on the InterPlanetary FileSystem (IPFS). This gives data scientists the necessary autonomy and ownership over the asset they create to be able to easily monetize their models without a third-party mediator.

% Move towards community-governed hubs. Goal is to achieve a censorship resistant asset sharing ecosystem

% decentralized hub of models and databases from any framework . 

% The current cloud infrastructure used by many individuals and companies in all areas of data science offer a plethora of tools tailored to most industry use cases of artificial intelligence. The advances of cloud storage and computing with services like AWS, Google Cloud, or Azure have accelerated AI development and allowed the large-scale adoption of machine learning. At the same time, the widely available credit grants for both individuals and startups have allowed early-stage ventures to bootstrap their AI products without significant financial overhead and deploy their models easily with the support of a global community.

% In Section \ref{s:dsworkflow}, we discuss the data science workflow. 
% Based on the needs of data scientists during this process
In Section \ref{s:existing}, we review existing AI Hubs. Based on our findings, we discuss some of the problems with existing hubs and associated libraries in Section \ref{s:problems}. Then, in Section \ref{s:infrastructure}, we discuss the potential advantages of decentralized technologies for these services. Finally in Section \ref{s:apps}, we present some of the hubs, libraries and frameworks that we have integrated with IPFS.

\section{Existing AI Hubs and Libraries} \label{s:existing}

The forms of existing AI Hubs have evolved as the requirements of AI and ML have become more clear. In this section, we review some existing AI Hubs such as GitHub, HuggingFace (HF) and ActiveLoop in terms of features and user base (as summarized in Table \ref{tab:hubs}). 

\begin{table}[!htp]\centering
\caption{Existing AI Hubs}\label{tab:hubs}
\scriptsize
\begin{tabular}{lrrrr}\toprule
\textbf{Existing AI Hub} &\textbf{GitHub} &\textbf{Huggingface}&\textbf{ActiveLoop} &\textbf{Replit} \\\midrule
\textbf{Launch} & 2008 & 2016& 2018 & 2016 \\
\textbf{Users} & SWEs & Data Scientists (NLP) & Data Scientists (CV) & SWEs \\
\textbf{IDE} & No & No & No & Yes \\
\textbf{Payments} & No & No & No & No \\
\textbf{Storage/Asset} & Code & Code, Datasets, Weights & Datasets & Code \\
\textbf{Compute/Hosting} & No & GPU (Inference) & No & CPU \\
\textbf{Cloud Infrastructure} & Centralized & Centralized  & Centralized & Centralized \\
\textbf{Governance} & Centralized & Centralized & Centralized & Centralized \\

\bottomrule
\end{tabular}
\end{table}

\subsection{GitHub}

GitHub\footnote{\url{https://github.com/}} is a platform for software development and version control that was launched in 2008. It builds on the open source git standard and adds a number of components for efficient collaboration, CI/CD, debugging, and unit testing. It also provides social networking functionality such as feeds, followers, and wikis. The site allows users to browse public repositories for software assets on the site. As of 2019, it had more than 40 million users and 100 million repositories (\cite{github2019state}).

GitHub has been heavily used for the development of AI and ML software in the form of tools, frameworks, and libraries. The number of AI and ML repositories increased slowly from 2009-2012 until an acceleration in 2017 (\cite{gonzalez2020state}). More applications of AI and ML are created annually than tools, libraries, and frameworks. It provides storage for software assets, but not large files such as datasets and model weights. The importance of these assets are an important differentiator between traditional software development and AI. To modify a code asset, a user employs a separate interactive code editor (IDE) and then pushes to GitHub through git at the command line. 

 % Interestingly, GitHub was originally a flat organization. 

% Interaction happens through git, no interactive code editor. 

% The application is built using Ruby on Rails, and it is hosted on Kubernetes clusters. 

% \subsection{Kaggle}

% Kaggle has originally started as an AI competition platform, on which data scientists compete to achieve the best performance on a particular dataset for a monetary prize. While this feature is still one of the reasons Kaggle gets so much user engagement, the hub has also expanded to include open-source datasets and educational materials on different topics in AI. Furthermore, Kaggle has developed a successful reputation scheme called the Progression System where users earn badges based on their activity in different parts of the platform.

% Owned by Google. Originally a competition platform. Kaggle datasets has been very successful. Can get paid. 

% Kaggle uses the Progression System for reputation https://www.kaggle.com/progression

\subsection{HuggingFace Hub}

HuggingFace (HF) is a company aiming to becoming the "GitHub for AI". HF achieved success by creating a unified library for the popular transformer models for natural language processing (NLP), making it easier to train, optimize, and deploy state-of-the-art model \cite{wolf2019huggingface}. They subsequently launched HF Hub\footnote{\url{https://huggingface.co/}}, a platform with features for code-sharing and collaboration such as discussions and pull requests (similar to GitHub). Unlike GitHub, HF also provide storage for large files (through git-lfs) such as datasets and pre-trained model weights (as well as code), meaning that ML developers can keep all of their assets in one place. Furthermore, users can use HuggingFace Spaces to host web-based demos of machine learning apps using the Gradio or Streamlit. They have built a large community of active data scientists, with the platform recently passing 100,000 users and 50,000 open source ML models.

To update an asset on HuggingFace Hub, data scientists must use a separate interactive code editor (IDE) with git through the command line. 

\subsection{ActiveLoop Hub}

Activeloop Hub\footnote{\url{https://www.activeloop.ai/}} is an open source library for efficiently storing and retrieving large machine learning datasets. Its data is stored in a columnar database management system, which is used to efficiently relate different files to each other to generate data samples via indexing. The core of the data layout is ActiveLoop’s chunking mechanism, which splits a dataset into so called chunks of data of size 16 MB, accelerating data streaming by sending more data in a single network request. The data layout allows efficient grouping of different parts of the dataset which is then automatically read by other services within the ActiveLoop ecosystem, such as the data visualizer.

\subsection{Replit}

Replit\footnote{\url{https://replit.com/}} is an online IDE. Unlike GitHub and HuggingFace, where modifying assets requires a separate IDE and command line, Replit users can interact with code and source control for their project through a web-based graphical user interface. Replit provides a shared compute engine that provides collaborative coding similar to Google Docs, where code can be run and displayed to multiple users. However, GPU support has not yet been released. Furthermore, file storage is limited to 0.5 GB for free users and 5 GB for paid users, which is too small for most ML assets. Replit has other features such as AI-assisted tools for software development, such as co-pilot and live chat and in-line threads for discussions around code by users.

\section{Problems with Existing AI Hubs and Libraries} \label{s:problems}

In the previous section, we explored the features of existing AI Hubs. With this information, we now analyze some of the issues with existing solutions.

\subsection{High Storage and Compute Costs} \label{s:cost}

The field of deep learning requires heavy amounts of storage and compute. Machine learning datasets often reach into the 100s of GBs, and pre-trained model weights can be large too. Furthermore, the computational cost of AI research is increasing exponentially with time, creating to higher barriers to entry for participants (\cite{schwartz2020green}). As a result, cloud services, such as storage and compute, are a significant expense for AI startups. Currently, three companies make up approximately two thirds of the market share of cloud service (\cite{cloudmarket}). More than half of Amazon’s profits has come from Amazon Web Services (AWS), and 20\% of AWS customers deliver 80\% of revenue with the widest margins come from small and medium-sized customers (\cite{amazon}). Popular AI Hubs like GitHub, HuggingFace, ActiveLoop and Replit rely exclusively on centralised cloud platforms.

\subsection{Lack of Monetization and Reward} \label{s:Ownership}

There are few online platforms where data scientists can get paid to work independently (\cite{marketplace}). GitHub Sponsors allows users to make monthly money donations to projects hosted on GitHub, but contributions are typically low. HuggingFace, ActiveLoop and Replit do not enable monetization by users. The payment infrastructure is primarily set up to transfer payments from the user to the platform itself. For example, while HuggingFace do offer free services and contribute to open source development, they also charge users for premium services that are not open source. In contrast, all contributions by users must be for free, with no ability to offer paid services. Data scientists often contribute to open source but still need to support themselves with income from elsewhere, such as jobs within tech companies or universities.

Open source tools and libraries are widely used by commercial platforms and products within software development and AI (\cite{langenkamp2022open}), although the contributions are not typically rewarded. Platforms invite assets to be uploaded by users, but do not share any generated revenue or platform ownership with users, even when directly monetizing their contributions. For example, GitHub Copilot is a commercial product for code generation that uses a model trained on user-contributed code. HuggingFace's paid inference API can be used to accelerate the deployment of user-contributed models. 

% OpenAI (who share a parent company with GitHub) trained the Codex model on user-contributed code. They subsequently subsequently commercialised the model despite much of the code having X license. Is this legal? Grey area. It is legal for a human to look at different samples of code and write their own from scratch. Similarly, while language models do train on this data, it could be argued that the models are interpolating between observed samples. Nonetheless, it shows that Microsoft are happy to act in an extractive manner with contributors to their platform.

\subsection{Lack of Control and Ownership} \label{s:Control}

% Users own the majority (79.1\%) of applied AI & ML repositories, but organizations own more (51.43\%) of the AI & ML tools (\cite{gonzalez2020state}).

% GitHub confirmed that it was now blocking developers in Iran, Crimea, Cuba, North Korea, and Syria from accessing private repositories

Generally, software developers and data scientists do not have full control and autonomy with their creations on centralized platforms. In one case, GitHub reverted malicious changes (and suspended the account) of a developer to their own popular open source library, raising questions around the rights of developers to do what they wish with their code \cite{developer}. In the field of AI, there has been an ongoing discussion on whether open sourcing disruptive models should be commonplace, since there is the potential for harm and bias. For example, AlphaFold can be used for discovery of novel toxic molecules. Language models can be trained on abusive content and used by online bots. Large models that are trained on the corpus of internet data reproduce bias within generated text and images. As a result, platforms like HuggingFace have come under pressure to gate or remove access to models.  On the other hand, it can be argued that open sourcing the model puts the technology in the hands of more people that can study and solve issues around safety and bias. In other words, there is an orthogonal risk involved with centralization of AI in the hands of a few. Keeping models closed source effectively turns large tech companies into gatekeepers, who may not always be relied upon to adjudicate on disputes in an unbiased manner. 

Finally, it is difficult for owners to manage fine-grained access to assets. It is common for data scientists to need to register for access to datasets online. After making a request to the owners (sometimes with information on the intended use case), the data scientist receives a link or a login. This could use a traditional access token like OAuth 2.0  (\cite{hardt2012oauth}) or API keys. This link or login for datasets and models can be widely shared, and licenses (e.g. restricting to non-commercial use) are often broken. While possible to keep repositories private, this is often a paid feature and the encryption key is held by the platform rather than the user.

% Users may still want to store the data locally. However, there are few options for the user to share an asset on a hub from their local machine. 

% Competitors in Kaggle competitions give away ownership of their solutions to the competition partner.  

% Finally, you don't even own your reputation on centralized hubs. 

% \begin{figure}[!t]
%     \centering
%     \includegraphics[width=0.5\linewidth]{figures/metahub.png}
%     \caption{The decentralized AI stack, consisting of the (i) Infrastructure Layer, (ii) Module Layer, and (iii) App Layer: contains a number of apps that make up MetaHub. The users of Metahub are the many stakeholders required to undertake successful projects.} 
%     \label{fig:idea}
% \end{figure}

\subsection{Difficult to Reproduce} \label{s:Reproduce}

The limitations of existing hubs such as GitHub for AI can make reproduciblity more difficult. For example, academic papers usually contain links that may include code on GitHub, and datasets and model weights stored on the cloud. Reproducing experiments is difficult and can require many steps such as downloading datasets, running processing scripts and installing environments, which is a time-consuming and tedious user experience. This issue results from a variety of factors such as the lack of standardisation and interoperability of in the format of assets (such as dataset and code), and the decoupling of assets from compute environments and infrastructure needed to operate on them. Some of these issues can be resolved by using containers and notebooks to replicate environments and bring compute to code. At the same time, notebooks can be difficult to deploy. HuggingFace Hub uses Gradio and Streamlit apps. Replit integrates code respotiories with compute environments, but has limited storage for assets such as datasets and model weights. 

\subsection{Lack of Standards}

Ideally, assets in an AI Hub should be modular, reusable and interoperable. The "interface" for the modules should have a common standard. However, this typically does not happen in practice. Datasets and model weights for deep learning are distributed across many websites and cloud platforms. There are few standards for the format in which a particular type of dataset is stored e.g. file structure, file names, file types. This makes joint training of models on multiple datasets more difficult. Furthermore, models themselves often have different formats. The success of HuggingFace is largely due to the success of the transformers library, which standardized the format for this popular family of models. 

% brain imaging dataset standard (BIDS) [cite] is a good example

% -----------------------------------------------------------------------------------------------------------------------------------------------------------------------------------------------------------------------------------------------------------------------------

% \section{Decentralized Infrastructure for AI Hubs} \label{s:daihub}

% Our solution for a decentalized AI Hub is composed of three layers, made up of infrastructure, modules and apps. This is shown in Figure \ref{fig:idea}. 

\section{Review of Decentralized Technologies for AI Hubs and Libraries} \label{s:infrastructure}

\begin{figure}[h]
    \centering
    \includegraphics[width=0.9\linewidth]{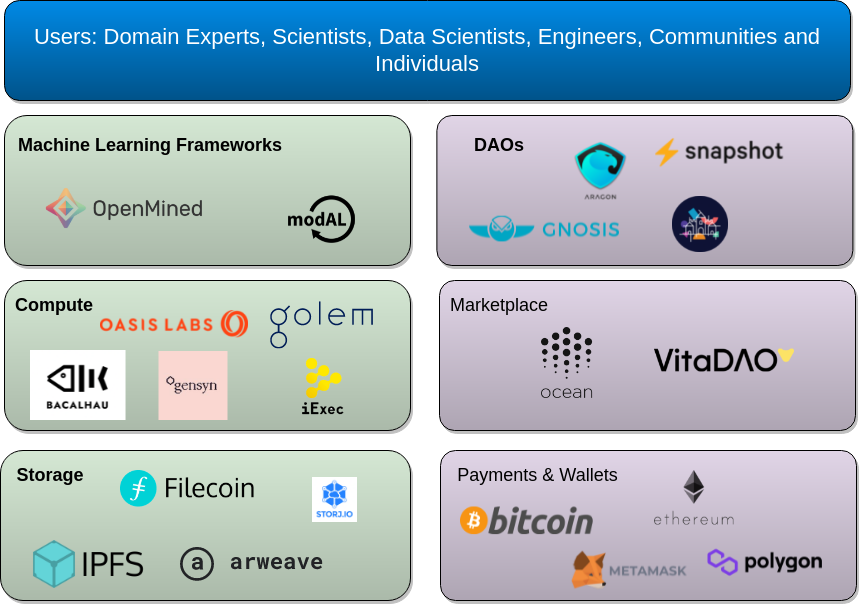}
    \caption{Decentralized infrastructure for AI Hubs, such as Web3 payments, wallets, marketplaces, storage and compute, learning frameworks and DAOs. The users of decentralized AI Hubs are the many stakeholders required for undertake successful projects.}
    \label{fig:metahub}
\end{figure}

In Section \ref{s:existing}, we discussed some of the features of existing AI Hubs. Decentralized technologies - such as Web3 payments, wallets, marketplaces, storage and compute, learning frameworks and DAOs - have the potential to alleviate some of the limitations of existing AI Hubs discussed above. Examples of projects working on these individual projects are shown in Figure \ref{fig:metahub}.

\subsection{Payments} 

There are few options for AI workers to monetize their creations and rewards generated by their contributions are often not shared, as discussed in Section \ref{s:Ownership}. We believe that building in mechanisms for monetization and ownership by users would create a healthier and more sustainable ecosystem and economy. This can be achieved using cryptocurrencies (such as Bitcoin, Ethereum, Polygon, Ocean and Filecoin) and stablecoins (such as DAI or USDC), which can be used for micro- and streaming payments to stakeholders such as data scientists, data providers and compute providers with low transaction fees. Thus, integrated payments offer many opportunities for use with machine learning frameworks such as active learning and data crowdsourcing.

% Previous hubs involve payments directly through their website, or through a specific cryptocurrency that they created. This limits developers into exchanging value for services based onto their own preferred assets. Allowing developers to monetize from their assets with any crypto currency is what a truly open hub strives to offer. A token can be issued for additional benefits, but should not be the sole means of transactions due to its aforementioned limitations. 

\subsection{Web3 Wallets}

As discussed in Section \ref{s:Control}, data scientists typically do not have control of what they create online. The platform is typically trusted as a middle man in control of your assets. Even if a repository containing assets is private, the platform holds the private keys. A Web3 wallet can be used to put the user in control of their private keys. The word wallet tends to have financial connotations. However, wallets are often used in the real world as a place where you hold ownership and identity documents (such as a driver’s license). Similarly, Web3 wallets can be used for ownership and identity in the digital world. Wallets are interoperable in the sense that you can use the same wallet to signify ownership of assets across many different protocols. It can also be used to replace a login. Web3 wallets include software wallets (such as MetaMask\footnote{https://metamask.io/}) and hardware wallets (such as Trezor\footnote{https://trezor.io/}).

\subsection{Marketplace} 

Traditional AI Hubs and marketplaces are typically operated by a centralized entity serving as a middle man. In the ideal scenario, the operator provides services in exchange for transaction fees and acts as a mediator for conflict resolution between users. However, centralized hubs and marketplaces also have the power to capture an outsized proportion of the value generated in a market economy as network effects grow. Furthermore, it is difficult to manage access to assets, and licenses for datasets and software are often broken. This contributes to the issues discussed in Sections \ref{s:Ownership} and \ref{s:Control}. 

Using decentralized marketplaces protocols for tracking publication, ownership of (and access to) assets has the potential to mitigate these risks. All operations are stored on an immutable public distributed ledger such that provenance can be tracked. For AI use cases, assets can include datasets, models, algorithms, apps, notebooks and manuscripts. Examples of decentralized marketplaces include Ocean Protocol (\cite{protocol2021tools}) and VitaDAO (\cite{golatowhitepaper}). These protocols use non-fungible tokens (NFTs) to represent ownership of the underlying intellectual property (IP), and fungible tokens to represent access rights to assets under different types of licenses. The details of published assets (and associated metadata) are encrypted and stored on-chain, along with access control parameters. A decentralized identifier (DID) is issued to represent the asset’s decentralized digital identity, and a DID Document (DDO) is used to include additional information relevant to the asset. This allows providers to include information relevant to the asset, and include additional fields for accommodating any asset that may need extra fields in its description.

Access gated by tokens on a blockchain has advantages compared to traditional access token like OAuth 2.0, by solving the "double spend problem". They act as access tokens that can only be used by one individual or for a period of time. If a user receives a token on a blockchain, the user can still share it with someone else but this means the original user will no longer have access. This facilitates more fine-grained access control by owners. 

% Smart contracts can also be used to automate adherence to service level agreements (SLAs). For example, payment returned if training on a dataset doesn't 

\subsection{Storage} 

While details about the assets are stored on-chain with decentralized marketplaces, the data associated with the asset are often too large to store on chain. As discussed in Section \ref{s:cost}, storage on centralized cloud providers is expensive. Furthermore, these services are less robust and more prone to censorship (see Section \ref{s:Control}). Popular dataset and model hubs like HuggingFace and ActiveLoop Hub rely on centralised cloud platforms. 

% Furthermore, users may want to store the asset locally while making it available to others. 

Decentralised protocols for storage have the potential to vastly reduce the costs incurred by data scientists for storing raw and processed versions of datasets and model weights. This makes  it possible to download files from multiple locations that aren't managed by a single organization.  The interplanetary file system (IPFS) (\cite{ipfs}) is a   peer-to-peer  protocol for storing and accessing data in a permissionless and censorship-resistant way. IPFS clusters enable data orchestration across swarms of IPFS peers by allocating, replicating, and tracking assets. Another important feature that IPFS offers is the ability to verify the validity of assets using Content Addressable Identifiers (CIDs), based on the content's cryptographic hash. This helps both builders and the consumers of data science products. A user can verify the model used is what was promised using the inbuilt cheksum method that the CID offer. 

% ML needs low latency

% Filecoin allows users to pay others for storage.

% Filecoin Virtual Machine (FVM) is a blockchain on top of IPFS. Potenti

\subsection{Compute} 

Access to compute is a necessity for AI projects, and the provision of services by a handful of centralized companies has resulted in inflated costs (see Section \ref{s:cost}). At the same time, the experiments and results of AI studies are often difficult to reproduce, as discussed in Section \ref{s:Reproduce}. While less mature than peer-to-peer storage solutions, decentralized protocols for providing compute resources aim to reduce the barrier-to-entry for compute providers and remove the centralised overheads on scaling (\cite{fieldingwhitepaper}). This provides more options for end consumers, resulting in reduced cost. Ideally, compute should be run where the data is stored - called Compute over Data (CoD) by the Bacalhau project\footnote{https://github.com/filecoin-project/bacalhau}, or Compute to Data (C2D) by Ocean Protocol - rather than transporting data to the location of the compute which is expensive. In this setting, decentralized compute infrastructure presents many opportunities for integration with privacy-preserving machine learning. With Ocean Protocol, users can act as compute providers. Compute providers may run a number of compute environments that may be specialized for different use cases and stages of development such as experimenting, training and deployment.

\subsection{Machine Learning Frameworks} 

Decentralizing infrastructure for storage and compute, and integrating payments has the potential to open up new use cases of AI. This require advancements in decentralized frameworks for machine learning. For example, privacy-preserving machine learning (PPML) - through libraries such as Openmined\footnote{https://www.openmined.org/} - has the potential to unlock learning on private data such as health records and user data. Integrated payments can be used with active learning frameworks with libraries (such as modAL, \cite{danka2018modal}) and tools for crowdsourcing human intelligence (such as Turkit, \cite{little2010turkit}). 

% Human in the loop (HITL)

\subsection{DAOs}  

Decentralized autonomous organizations (DAOs) are systems that allow communities to coordinate and take part in self-governance, as determined by a set of self-executing rules on a blockchain (\cite{hassan2021decentralized}). DAOs have previously been suggested as governance structures for digital data trusts (\cite{nabben2021decentralised}). DAOs are sometimes imagined as being governed by autonomous algorithms, with humans at the margins. However, there is an increasing push towards a future of collective intelligence that promotes harmony between humans and algorithms by optimizing for the autonomy of individuals (\cite{nabben2021imagining}). Currently, most DAOs are superficially decentralized, with most involving high concentration of voting power by a select few. 

% How ownership of assets within VitaDAO works. FRENS and FAM (watch DeSci Berlin video). 

We suggest that DAOs can be used to (i) govern assets within AI Hubs, and (ii) to create decentralized AI Hubs that are governed by communities rather than single entities. Tools for governing assets within DAOs include multisignature wallets (such as Gnosis\footnote{https://gnosis-safe.io/}) and profit-sharing mechanisms (such as Superfluid\footnote{https://www.superfluid.finance/}). Multisignature wallets provide functionality for sharing ownership and control of assets with multiple individuals in teams in a trustless manner, while profit-sharing mechanisms can be used to distribute the revenue generated by assets. Tools for governing the infrastructure of AI Hubs include decentralized voting systems (such as Snapshot\footnote{https://snapshot.org/}).

\section{Integrations of IPFS with AI Hubs, Libraries and Frameworks} \label{s:apps} 

AI Hubs typically contain many features, as discussed in Section \ref{s:existing}. Reproducing this functionality using decentralized technologies (as discussed in the previous section) is a large undertaking. In this section, we discuss the implemented solutions to date in the form of Python libraries, a web app and notebook environment. 

% The apps correspond to different stages of the data science workflow.

% The first few stages happen in notebooks. Need Python libraries to interact with the Metahub. IPFS acts like a FUSE mount. When the steps are refined, we can begin to use pipelines, before deploying our model. 

\subsection{Python Libraries for IPFS} 

Decentralised storage solutions have the potential to vastly reduce the costs incurred by data scientists for storing raw and processed versions of datasets, as well as model weights. Currently, there are few tools for interacting with decentralised storage in the language that data scientists are used to i.e. Python. This is especially true for writing to storage. Furthermore, IPFS is not well integrated with AI Hubs and libraries.

% In the previous sections, we learned about the data science workflow. Data scientists need to interact with storage. Decentralized storage solutions, like IPFS, are not commonly used in data science. We believe this could be the result because of the mismatch in the programming language of choice between IPFS (and its APIs) developed in GoLang and JavaScript, and the ones used by data scientists, namely Python, Matlab, or Julia.

% We are working towards addressing this by creating Python libraries for interacting with IPFS.

\subsubsection{IPFSSpec (and HuggingFace)} 

\begin{figure}[!t]
    \centering
    \includegraphics[width=1\linewidth]{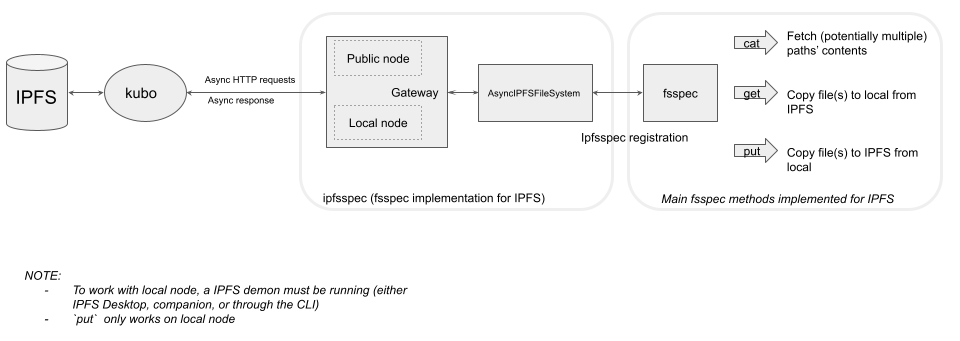}
    \caption{Schematic of ipfsspec.} 
    \label{fig:ipfsspec}
\end{figure}

Filesystem Spec (fsspec) is a project to provide a unified pythonic interface to local, remote and embedded file systems and bytes storage. It is used by HuggingFace, Pandas, and Dask. Previously, a solution for a read-only version of fsspec (called ipfsspec) existed. We implemented write functionality for ipfsspec\footnote{\url{https://github.com/AlgoveraAI/ipfsspec}}, and re-implemented much of the read functionality. By doing so, read and write to IPFS is now supported by dependent libraries such as HuggingFace, Pandas, and Dask. The architecture of ipfsspec is shown in Figure \ref{fig:ipfsspec}.

% intended to be a place to define a standard interface that such file-systems should adhere to, such that code using them should not have to know the details of the implementation in order to operate on any of a number of backends. With hope, the community can come together to define an interface that is the best for the highest number of users, and having the specification makes developing other file-system implementations simpler. File system abstraction. Use where a global file system is needed. E.g. data scientists collaborating in shared notebooks.

% The goal of IPFSspec is to seamlessly integrate with widely used data science packages and storage providers in the data science community. A standard library to send and load data in Python is through FSspec. FSspec is the file system for storage providers such as Hugging Face, Google Drive, AWS S3, and Anaconda and is supported in data science python libraries such as Pandas and Dask.  Using FSspec as the foundation for the IPFSpy implementation accelerates the adoption as these platforms and libraries have large communities.  The functionality includes open, put, and get files; more details are found in the diagram below.

\subsubsection{IPFSPy} 

\begin{figure}[!t]
    \centering
    \includegraphics[width=1\linewidth]{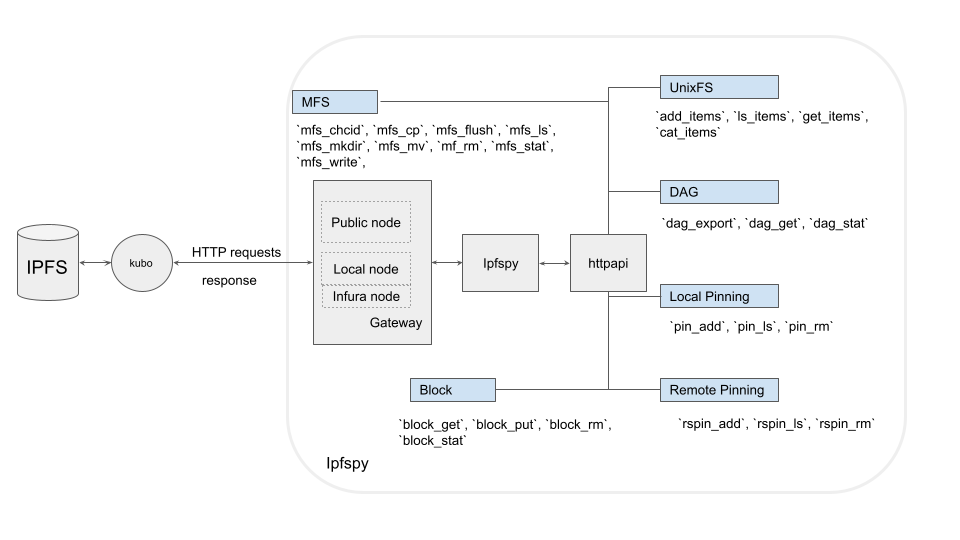}
    \caption{Schematic of ipfspy.} 
    \label{fig:ipfspy}
\end{figure}

While ipfsspec has the advantage of being used by other data science libraries, it supports a limited set of functions such as cp, rm, cat and mkdir. The ipfspy\footnote{\url{https://github.com/AlgoveraAI/ipfspy}} library provides a thin wrapper around endpoints for the IPFS HTTP API, as well as Estuary and Pinata APIs. It incorporates the ipfsspec library while also providing the Python community with a wider set of functionalities for interacting with the various components of IPFS and Filecoin, such as MFS, UnixFS, DAG, Local Pinning, Remote Pinning, and Block uses. This makes it easier to build custom solutions on top of IPLD for data storage and loading within specific AI and ML use cases. At the same time, we think that re-implementing the functionality of the various building blocks of IPFS in Python (i.e. without using the IPFS HTTP API) would further improve customizability. The architecture of ipfspy is shown in Figure \ref{fig:ipfspy}.

\subsection{IPFS on ActiveLoop}

Activeloop Hub is a repository for machine learning datasets and a library for efficient data streaming. The Hub introduces a data chunking mechanism extending the Zarr standard to be able to train machine learning models faster without requiring the data scientist to download large datasets locally. To enable the use of data streaming for the Web3 data science community, we integrated ipfpy within Activeloop Hub\footnote{\url{https://github.com/AlgoveraAI/Hub}} to enable decentralized storage for ActiveLoop Hub datasets. The integration is fully interoperable with the existing Hub library and allows the user to select any IPFS Gateway supporting read/write functionality.

\subsection{IPFS on Ocean Marketplace}

The Ocean Marketplace\footnote{\url{https://market.oceanprotocol.com/}} is an open source community marketplace for data, built on Ocean Protocol. It allows data providers to publish and monetize data assets. Most users store their assets on centralized storage such as Google Drive. As part of core tech grants for Ocean, we integrated storage using IPFS and Filecoin. 

\subsection{IPFS on Metahub}

There are several existing Web2 AI Hubs for datasets, models, apps and other assets, such as HuggingFace Hub and ActiveLoop Hub. However, these platforms are centralized with limitations discussed in Section \ref{s:existing}. IPFS and Filecoin allow users to store and retrieve assets using a peer-to-peer, rather than client-server, model. However, the discoverability of assets stored on IPFS is an issue. The Ocean Protocol marketplace facilitates ownership, monetization and discoverability of assets, although it has not been tightly integrated with storage solutions. Metahub\footnote{\url{https://metahub.algovera.ai/}} is our implementation of an AI Hub that combines the best parts of IPFS and Ocean Protocol and integrates with HuggingFace and ActiveLoop Hub, to create a Web3 AI marketplace where data scientists can use datasets and generate revenue from the algorithms that they develop. We also wrote scripts for scraping HuggingFace and ActiveLoop Hub datasets, downloading them, uploading to IPFS and publishing to the marketplace. 

\subsection{IPFS on Streamlit} 

Streamlit\footnote{\url{https://streamlit.io/}} is an open source library for building interactive AI apps in Python. Its main use case is allowing data scientists to quickly build and share demos of their ML models and it has been used extensively in open source AI communities. Streamlit is also integrated into HuggingFace Hub. However, Streamlit did not previously have support for Web3 functionality. We have integrated the Streamlit app framework with MetaMask\footnote{\url{https://github.com/AlgoveraAI/streamlit-metamask}}, IPFS/Filecoin and Ocean Protocol, meaning that data science apps can now build on these Web3 components. For example, users can now interact with components (like buttons) to log in with their Web3 wallet, store assets on IPFS and run compute-over-data directly from Streamlit apps. In future, we plan to embed Web3-integrated Streamlit apps directly into Metahub. 

% \subsubsection{metahub.py}

% \subsection{Data Labelling} 

% Some sort of app that provides and interface for humans to label data, that is then added to Metahub.

\subsection{IPFS in Custom Notebook Environments} \label{s:notebook}

Notebooks are one of the primary environments where data scientists perform exploratory data analysis (EDA) and build proof-of-concept workflows. The ipfsspec and ipfspy libaries can be used within notebook environments for interfacing with decentralized storage. In addition, we have created Jupyter Lab extensions\footnote{\url{https://github.com/AlgoveraAI/jupyterlab_extensions}} to log in with Metamask, upload data to IPFS/Filecoin and publish assets to the Ocean marketplace through the frontend. 

% \begin{figure}[!t]
%     \centering
%     \includegraphics[width=1\linewidth]{figures/workflow.png}
%     \caption{. \RB{Arshy, can you upload a HR version of this image? Also maybe we could slightly modify the image based on interactions with Metahub instead of storage?}} 
%     \label{fig:workflow}
% \end{figure}

% \subsection{Pipelines} 

% \subsection{Deployments} 

% Now the model is production ready. The model is deployed and its performance and other performance-related data are gathered to study the model’s performance in the wild.  

\section{Conclusion}\label{s:conclusion}

In this report, we (i) reviewed existing hubs and libraries for AI development, (ii) discussed some of the problems with existing solutions, (iii) discussed the potential advantages of decentralized technologies for these services, and finally (iv) presented some of the hubs, libraries and frameworks that we have integrated with IPFS.

\if0\blind{
\section*{Acknowledgements}
The authors acknowledge the generous support from the Filecoin Foundation (devgrant \#517) and Ocean Protocol Foundation.	} \fi

\bibliographystyle{chicago}
\spacingset{1}
\bibliography{IISE-Trans}
	
\end{document}